\documentclass[final,5p,times,twocolumn]{elsarticle} 

\journal{Nuclear Instruments and Methods A}

\usepackage{amsmath}
\usepackage{caption}
\usepackage{hyperref}

\begin{document}

\begin{frontmatter}

\title{Expected performance of the ALPIDE pixel layers in ALICE FoCal}

\author[a]{Jie Yi\corref{cor1} \corref{speaker}}
\cortext[speaker]{Speaker}
\author[b]{Max Philip Rauch}
\author{on behalf of the ALICE FoCal Collaboration}

\address[a]{Institute of Particle Physics, Central China Normal University, Wuhan, China}
\address[b]{Department of Physics and Technology, University of Bergen, Bergen, Norway}

\begin{abstract}
The ALICE experiment is designed to study ultra-relativistic heavy-ion collisions at the  Large Hadron Collider (LHC). As part of its major upgrades for Run 4, the Forward Calorimeter (FoCal) will be installed during Long Shutdown 3 (LS3). FoCal enables precise measurements of direct photon production at forward rapidity, providing a sensitive probe of the gluon distribution in protons and nuclei. 

This paper introduces the expected performance of forward electromagnetic calorimeter (FoCal-E) and presents several potential strategies for mitigating occupancy and BUSY violation challenges in the pixel layers of FoCal-E. Beam test results demonstrate that back biasing effectively reduces pixel occupancy. Meanwhile, SystemC simulations explore additional mitigation strategies—such as grid masking and a regional trigger—to further minimize BUSY violations and enhance detector performance under high-luminosity conditions.

\end{abstract}


\end{frontmatter}


\section{Introduction}

The A Large Ion Collider Experiment (ALICE) at the LHC investigates strongly interacting matter under extreme energy densities, focusing on the properties of quark-gluon plasma (QGP) formed in heavy-ion collisions. A high-precision Forward Calorimeter (FoCal) is under development for Run 4 to study the small-$x$ gluon structure of protons and nuclei by measuring direct photons, neutral hadrons, jets, and their correlations at forward rapidity (3.4 $< \eta <$ 5.5), as well as J/$\psi$ production in ultra-peripheral heavy-ion collisions \cite{ALICE-PUBLIC-2023-001}.

\begin{figure}[hb]
    \centering
    \begin{minipage}{\linewidth}
        \centering
        \includegraphics[width=0.65\linewidth]{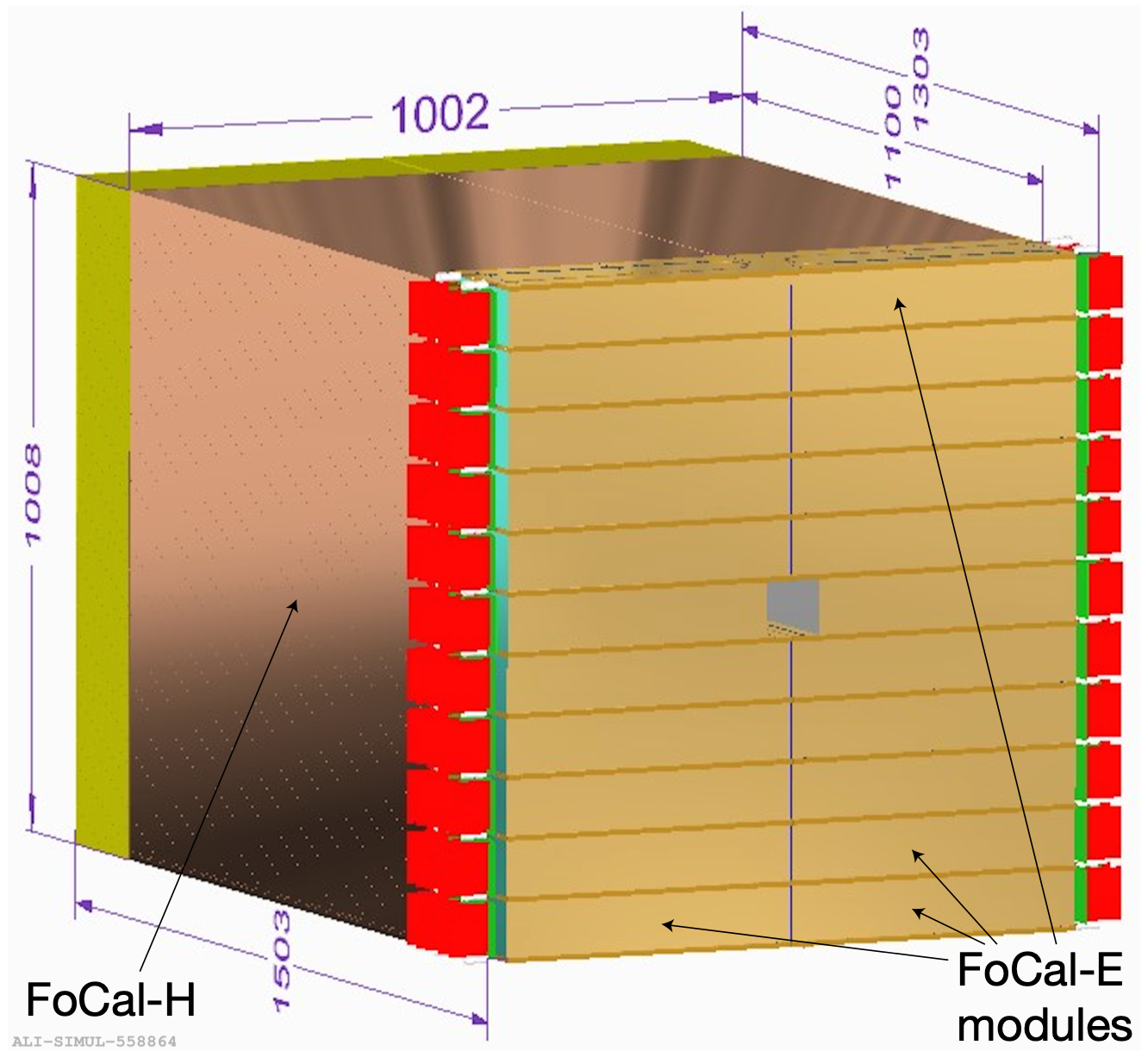}
        \caption{Schematic of FoCal, including FoCal-E and its central opening (84×84 mm²) for the beam pipe. Dimensions in mm \cite{CERN-LHCC-2024-004}.}
        \label{fig:FoCal}
    \end{minipage}
\end{figure}

FoCal has two main components (\autoref{fig:FoCal}): a highly granular electromagnetic calorimeter (FoCal-E), consisting of two pixel layers and 18 silicon pad layers, and a hadronic calorimeter (FoCal-H) of metal-scintillator design. Achieving fine spatial resolution for $\gamma - \pi^0$ discrimination motivates the selection of the ALICE Pixel Detector (ALPIDE), originally developed for the ALICE Inner Tracking System (ITS2), for the pixel layers.

However, operating ALPIDE sensors in a high-luminosity, high-occupancy environment poses challenges. High occupancy can trigger BUSY violations (when internal ALPIDE buffers are filled), impairing data integrity and trigger efficiency. Developing mitigation strategies is thus essential for ensuring FoCal-E’s performance under LHC Run 4 conditions. This work investigates occupancy and BUSY violation in FoCal-E pixel layers, using SystemC simulations and beam test data.

\section{The Pixel Layers of FoCal-E}

For FoCal-E, a small transverse shower size is crucial to optimize photon separation and reduce occupancy. 
Tungsten (W) was selected for its small Molière radius ($R_M=0.9$\,cm) and short radiation length ($X_0=3.5$\,mm), 
allowing a compact geometry with a sampling thickness of $\sim1\,X_0$, which is sufficient given FoCal-E’s moderate energy resolution requirements. 
Additionally, this compact design minimizes lateral shower spread, aiding in the discrimination of overlapping photon showers in high-multiplicity environments. 
While a conventional calorimeter design might set the transverse cell size near $R_M$, simulations indicate that granularity of $\sim0.3$\,mm significantly aids $\pi^0$ identification \cite{CERN-LHCC-2020-009}, improving electromagnetic shower reconstruction and enabling two-photon separation from $\pi^0$ decays at about 1\,mm. 
This finer granularity ultimately enhances direct photon identification.

\begin{figure}[h]
    \centering
    \begin{minipage}{\linewidth}
        \centering
        \includegraphics[width=0.7\linewidth]{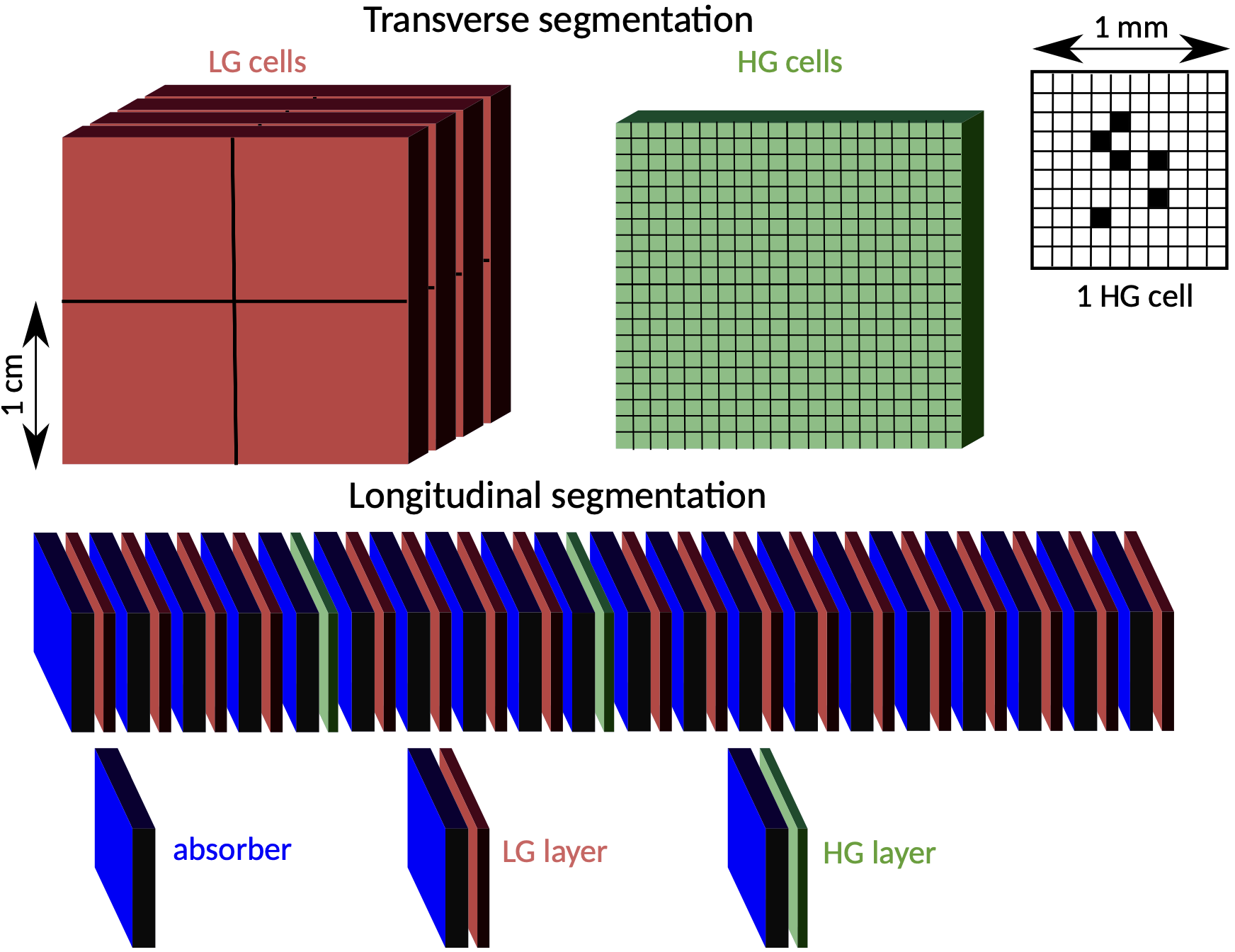}
        \caption{Sketch of FoCal-E, showing 20 Si-W layers: 18 pad sensor layers and 2 pixel sensor layers at positions 5 and 10, each individually read out \cite{CERN-LHCC-2024-004}.}
        \label{fig:FoCal-E-seg}
    \end{minipage}
\end{figure}

In \autoref{fig:FoCal-E-seg}, the high-granularity pixel layers of FoCal-E are shown at layers 5 and 10, utilizing the ALPIDE. The remaining 18 layers consist of low-granularity pad layers with a transverse cell size of 1 cm$^2$, read out and digitized by the High Granularity Calorimeter ReadOut Chip (HGCROC) for silicon pad sensors. The blue regions in the figure represent the absorber material, where each layer consists of a tungsten (W) plate with a thickness of approximately 1$X_0$, followed by an active silicon sensor-based structure.

\begin{figure}[h] 
\centering 
\includegraphics[width=\linewidth]{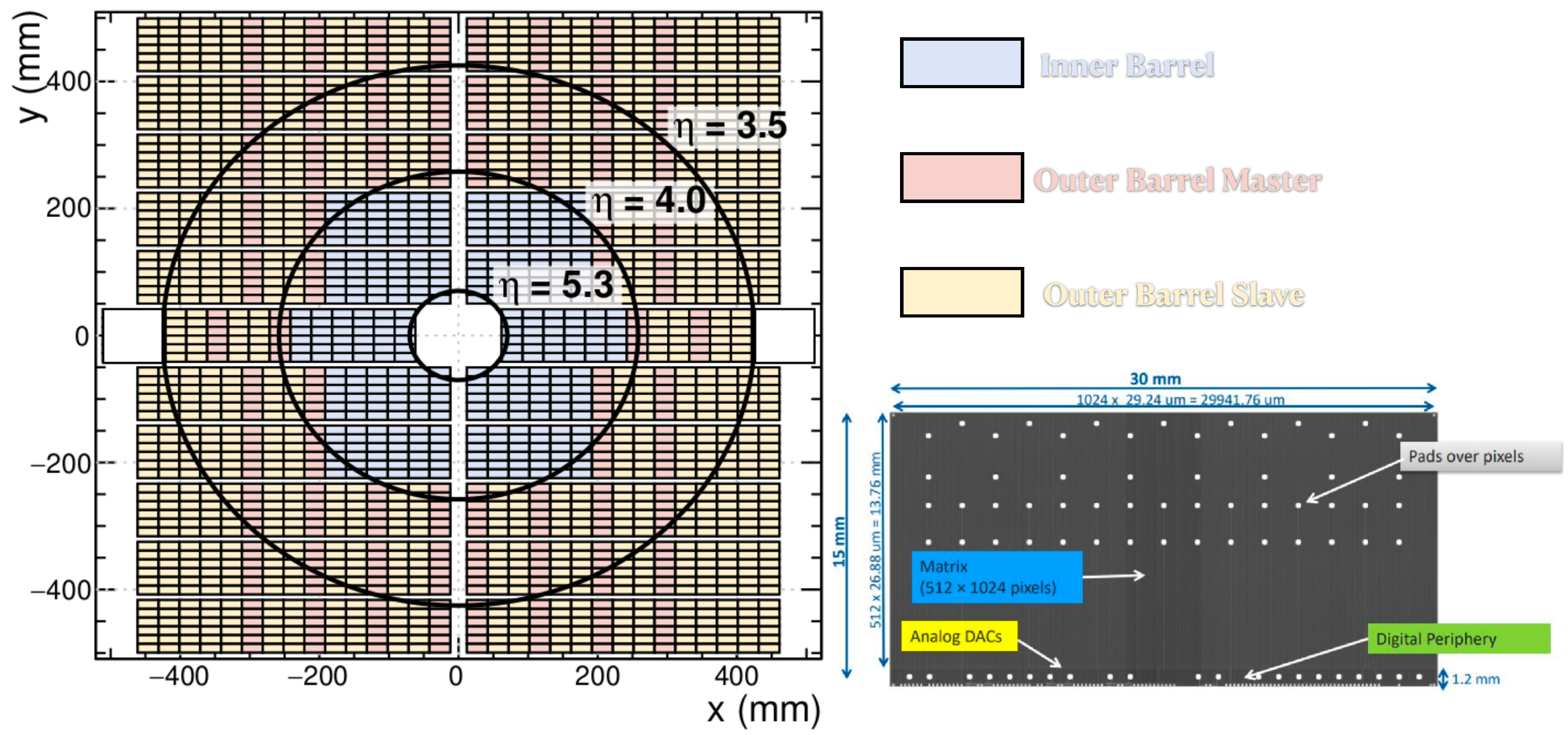} 
\caption{(Left) FoCal-E pixel layer plane, showing different ALPIDE sensor modes: IB (Inner Barrel), OB (Outer Barrel) slave, and OB master. (Right-bottom) ALPIDE layout showing key component positions \cite{CERN-LHCC-2024-004}.}
\label{FoCal_Pixel_Plane} 
\end{figure}

Each pixel layer comprises 22 patches (11 per side), each formed by 6 ALPIDE strings, for a total of 3,888 ALPIDE sensors. At the high-occupancy center, sensors operate in IB mode (960\,Mbps), while OB master/slave modes (320\,Mbps shared link) cover the periphery (\autoref{FoCal_Pixel_Plane}). Each ALPIDE measures 30\,mm~$\times$~15\,mm and contains a 1024\,$\times$\,512 pixel matrix, with digital readout circuitry along its lower edge. Every 29.24\,$\mu$m~$\times$~26.88\,$\mu$m pixel consists of a charge collection diode, amplification circuit, and discriminator.

\section{BUSY Violation}

Each ALPIDE contains a Multi-Event Buffer (MEB) with three slots for storing hit information. During operation, the ALPIDE opens a strobe window, and if a showering particle hits a pixel during this window and exceeds the threshold, the hit is stored in an available slot for later readout. However, in regions with high local occupancy, the rate at which hits are stored can exceed the readout speed. If all three slots in the MEB are filled before any slot is freed, a BUSY violation is triggered, preventing the chip from accepting additional hits until at least one slot is cleared. The period during which the chip cannot register hits is referred to as dead frames.

Minimizing BUSY violation in the central region is crucial for optimizing detector performance, ensuring efficient hit recording and reducing dead time in high-occupancy environments.

\section{Expected Performance in SystemC simulation}
To evaluate the performance of the pixel layers under high-luminosity conditions, a dedicated simulation framework was developed using SystemC, which was originally developed for ITS2 \cite{Nesbo2022}. This framework enables a detailed digital simulation of the ALPIDE-based readout chain, allowing the study of key performance parameters such as occupancy, BUSY violation rates, and data throughput. 

\begin{figure}[h]
    \centering
    \includegraphics[width=0.9\linewidth]{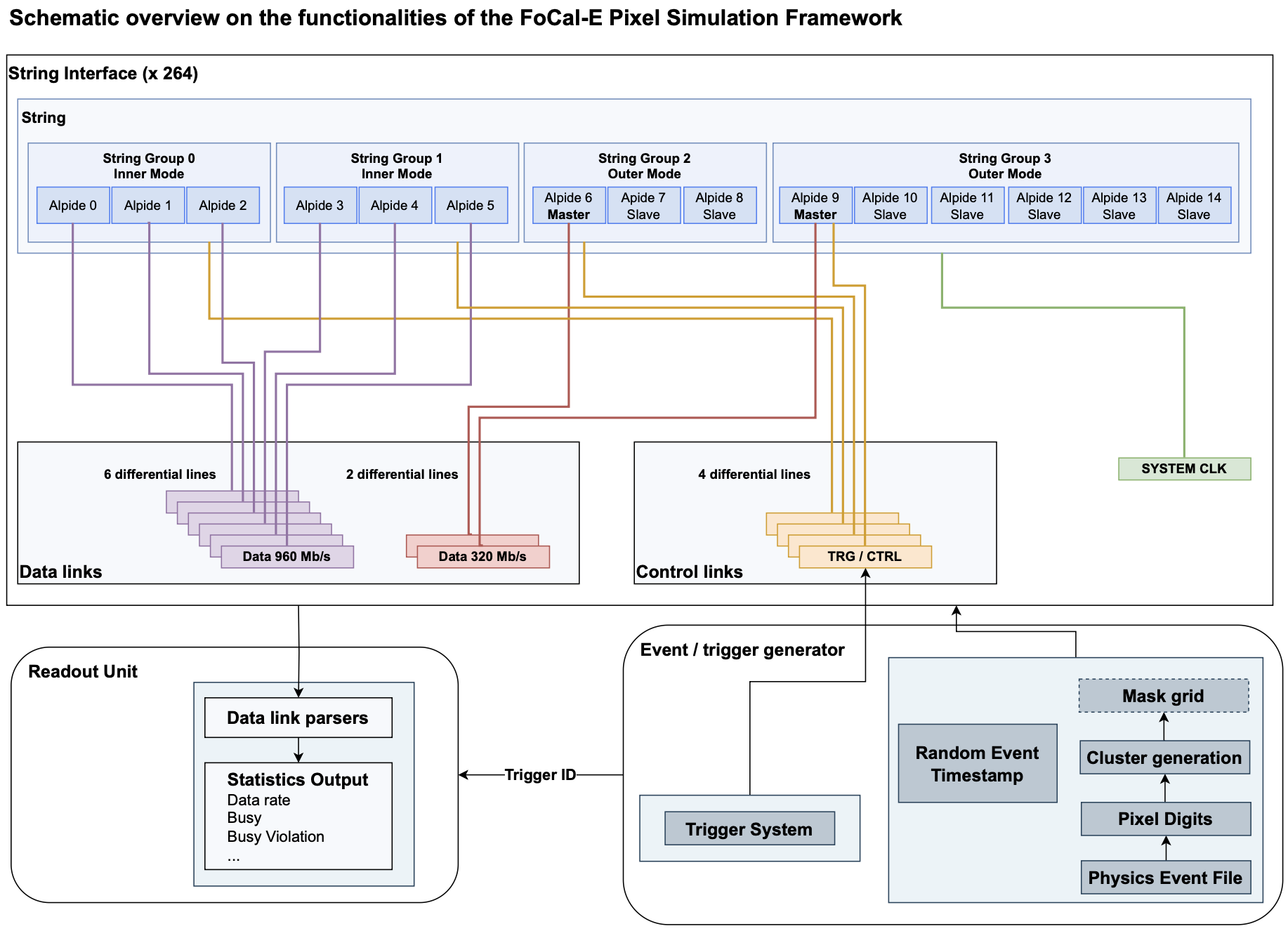}
    \caption{Overview of the SystemC simulation framework for FoCal-E pixel layers. The model simulates 264 pixel strings with 15 ALPIDEs each and includes a statistical readout analysis module\cite{CERN-LHCC-2024-004}.}
    \label{SystemC_Framework}
\end{figure}

\autoref{SystemC_Framework} illustrates the simulation framework used to model the FoCal-E pixel layers. It processes simulated minimum-bias events from PYTHIA, incorporating timestamp assignments and pixel-cluster generation for realistic occupancy modeling. 

The string-interface class handles input/output, and an event-generator class timestamps interactions using a probability model with a default interaction frequency of 1 MHz. Particle hits are mapped into the ALPIDE coordinate system, with a cluster-generation model producing Gaussian-distributed clusters (mean size $n_{\text{cl.size}}$ = 4 pixels, and the width $\sigma_{\text{cl.size}}$ = 1.4), refined using beam test validation. The trigger-generator class supports both periodic triggering, with a default frequency of f\textsubscript{trg} = 100 kHz, and externally derived triggers, allowing potential integration with the Fast Interaction Trigger (FIT) or a FoCal self-triggering scheme. To ensure realistic conditions under high-luminosity operation, the model accounts for the ALPIDE active time window (5 µs), simulating pixel activity and data-link readout with the bunch crossing timestamp (25 ns) granularity.

\begin{figure}[h]
    \centering
    \includegraphics[width=0.9\linewidth]{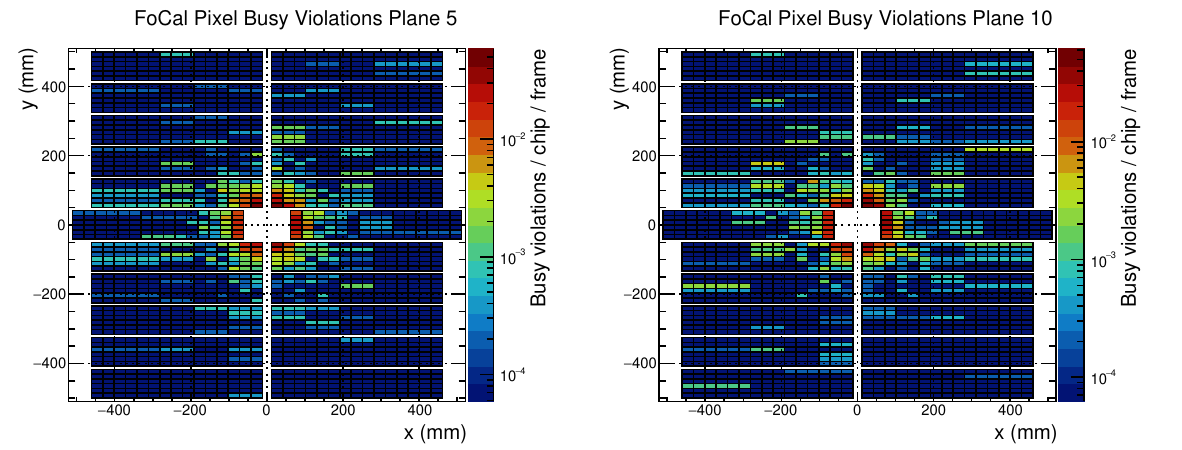}
    \caption{BUSY violation rate per ALPIDE in pp-collisions}
    \label{fig:ABUSYV}
\end{figure}

Based on the default simulation settings, the pixel layer was modeled under periodic triggering, and the resulting distribution of BUSY violation rate across the pixel layer is shown in \autoref{fig:ABUSYV}. The highest BUSY violation occurs in chips surrounding the beam pipe, reaching approximately 6\%, while it rapidly decreases with increasing radial distance.

\begin{figure}[h]
    \centering
    \includegraphics[width=0.9\linewidth]{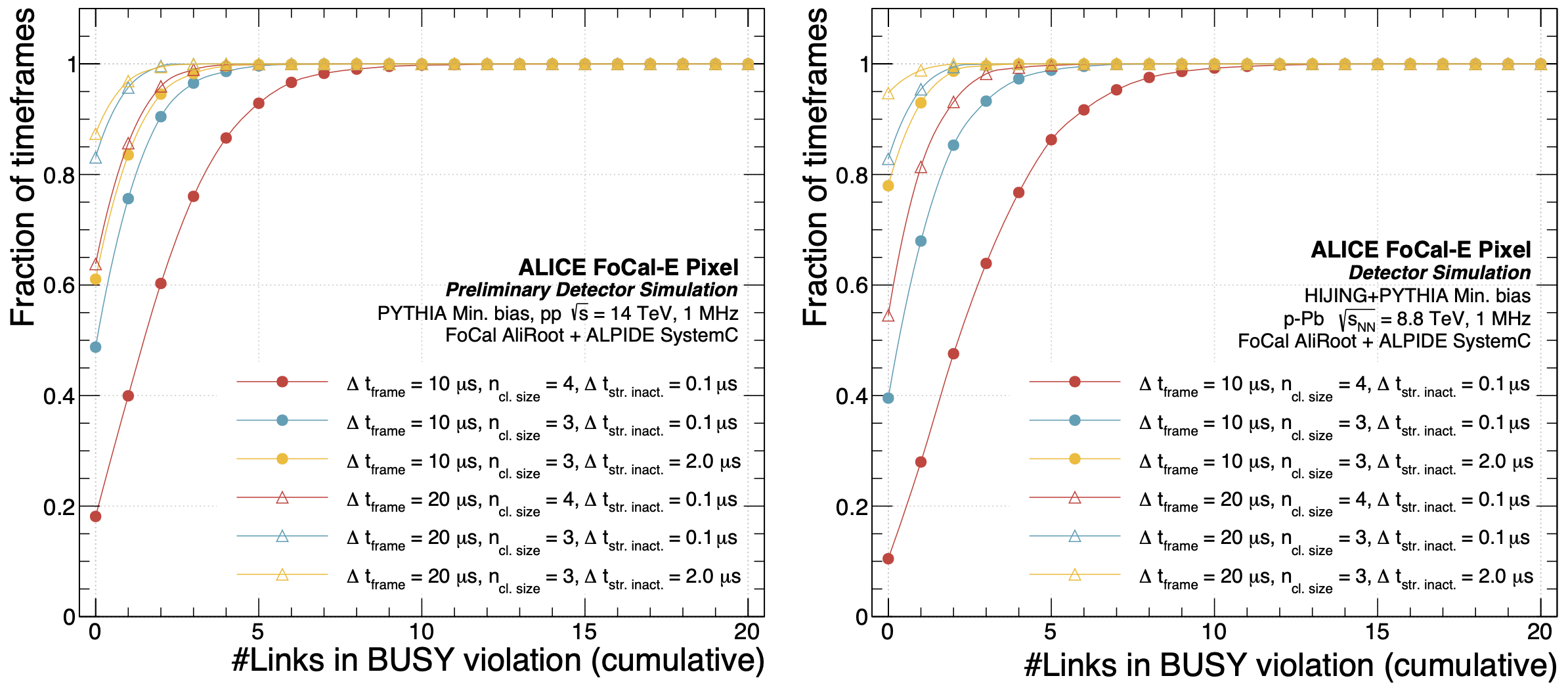}
    \caption{The fraction of frames with multiple links in BUSY violation \cite{CERN-LHCC-2024-004}.}
    \label{fig:BUSYV.fraction.of.timeframes}
\end{figure}

Various measures are investigated to reduce frame loss from BUSY violations, including back biasing ALPIDEs (lowering the mean cluster size from 4 to 3 and thus reducing the occupancy by 25\%), extending the time-frame length \(\Delta t_{\mathrm{frame}}\), and increasing the strobe-inactive time \(\Delta t_{\mathrm{strobe,inactive}}\). Lowering the cluster size and lengthening \(\Delta t_{\mathrm{strobe,inactive}}\) substantially raise the fraction of frames unaffected by BUSY violations. As shown in \autoref{fig:BUSYV.fraction.of.timeframes}, combining these settings can achieve up to \(\sim 87\%\) of frames with no violations, and \(\sim 97\%\) with at most one. Although a longer \(\Delta t_{\mathrm{frame}}\) further improves efficiency, it may increase pileup and chip occupancy, requiring a balance between readout margin and event overlap.

\begin{figure}[hb!t]
    \centering
    \begin{minipage}{0.86\linewidth}
        \includegraphics[width=\linewidth]{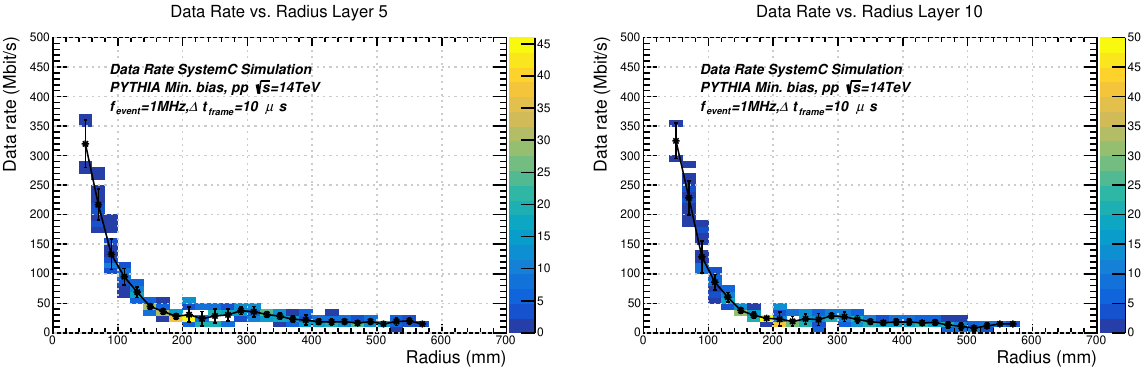}
    \end{minipage}
    \begin{minipage}{0.86\linewidth}
        \includegraphics[width=\linewidth]{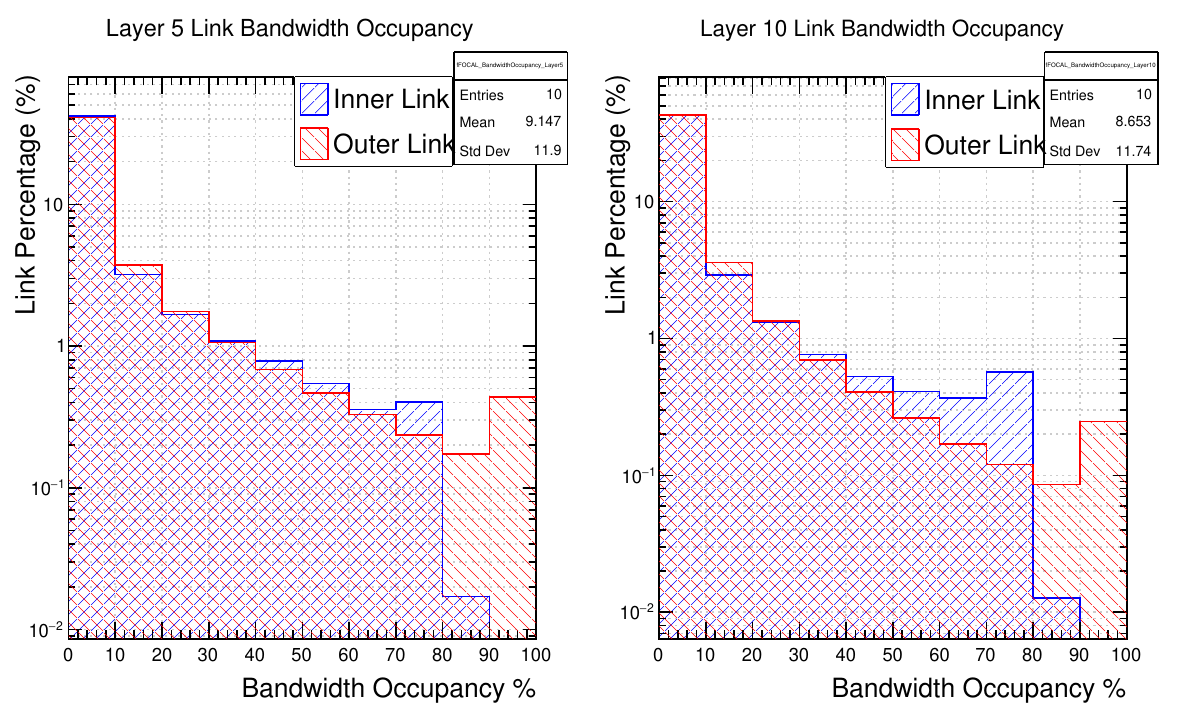}
    \end{minipage}
    \caption{(Top) Average data rate and (bottom) instantaneous bandwidth occupancy distribution with default setting.}
    \label{data_rate}
\end{figure}

Additionally, the SystemC simulation evaluated ALPIDE readout under both average and instantaneous data rates. As shown in \autoref{data_rate}, the IB and OB data links maintain a comfortable bandwidth margin when considering the average data rate. Under instantaneous conditions, IB links still preserve at least a 10\% bandwidth margin, whereas OB links occasionally approach saturation.

\section{Beam Test Results}

Four FoCal-E pixel layers were tested at the CERN SPS H2 beamline in September 2024, with 
ALPIDEs operated at back-bias voltages from 0\,V to 5\,V. The primary objective was to 
evaluate the reduction of pixel-layer occupancy in high-energy electromagnetic showers under 
various back-bias settings.
\begin{figure}[!h]
    \centering
    \begin{minipage}{0.43\linewidth}
        \centering
        \includegraphics[width=\textwidth]{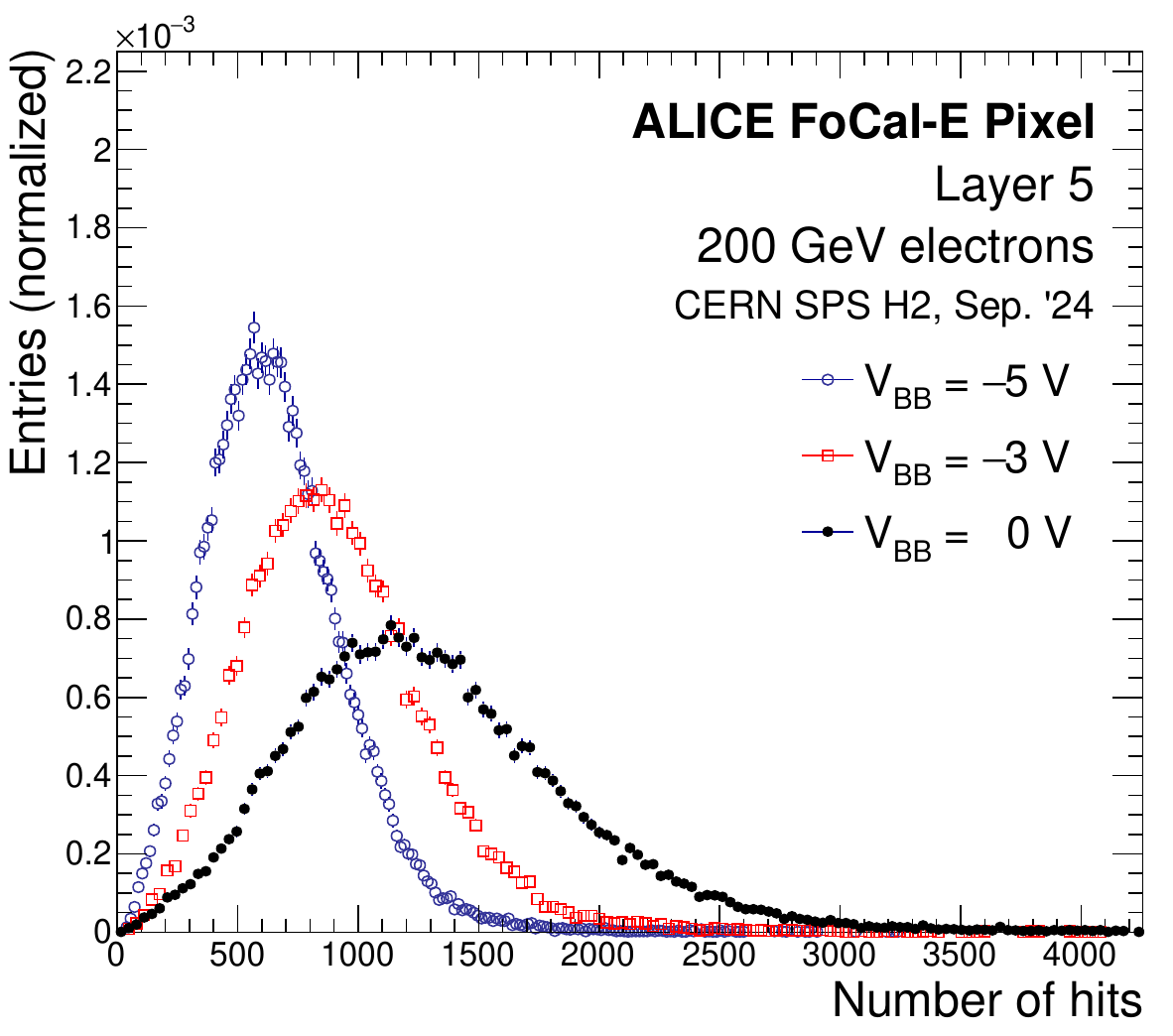}
    \end{minipage}
    \begin{minipage}{0.43\linewidth}
        \centering
        \includegraphics[width=\textwidth]{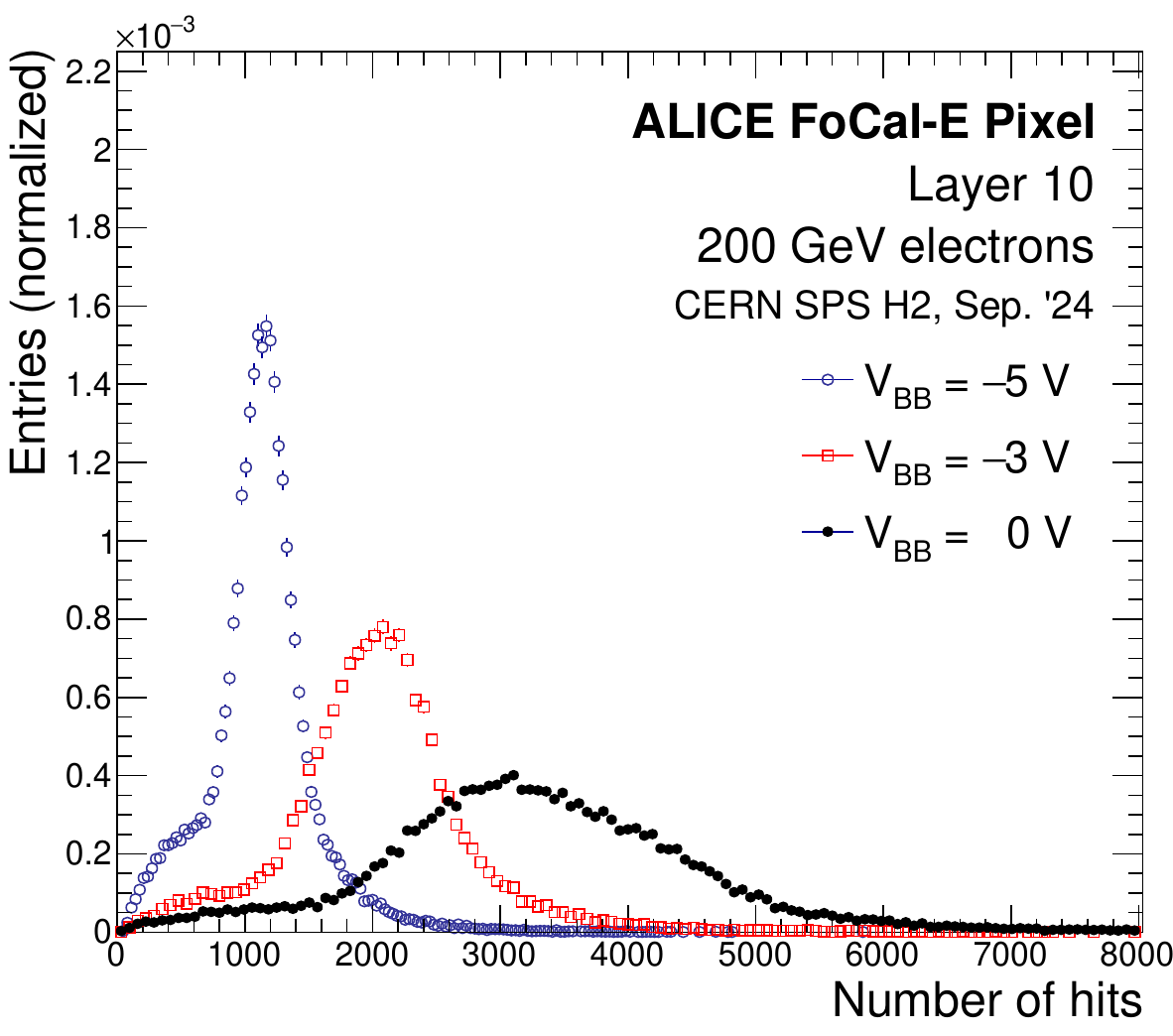}
    \end{minipage}
    \begin{minipage}{0.43\linewidth}
        \centering
        \includegraphics[width=\textwidth]{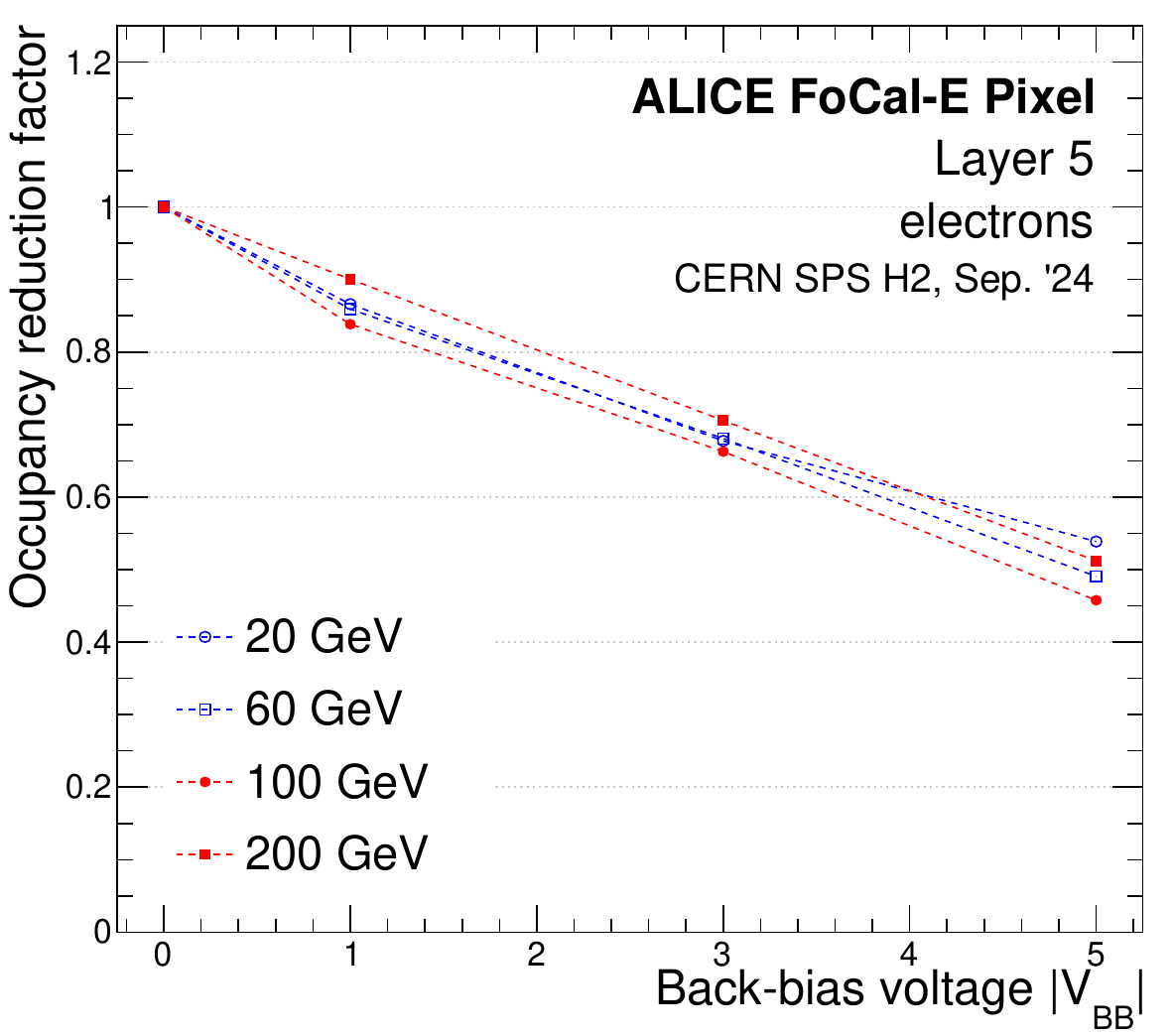}
    \end{minipage}
    \begin{minipage}{0.43\linewidth}
        \centering
        \includegraphics[width=\textwidth]{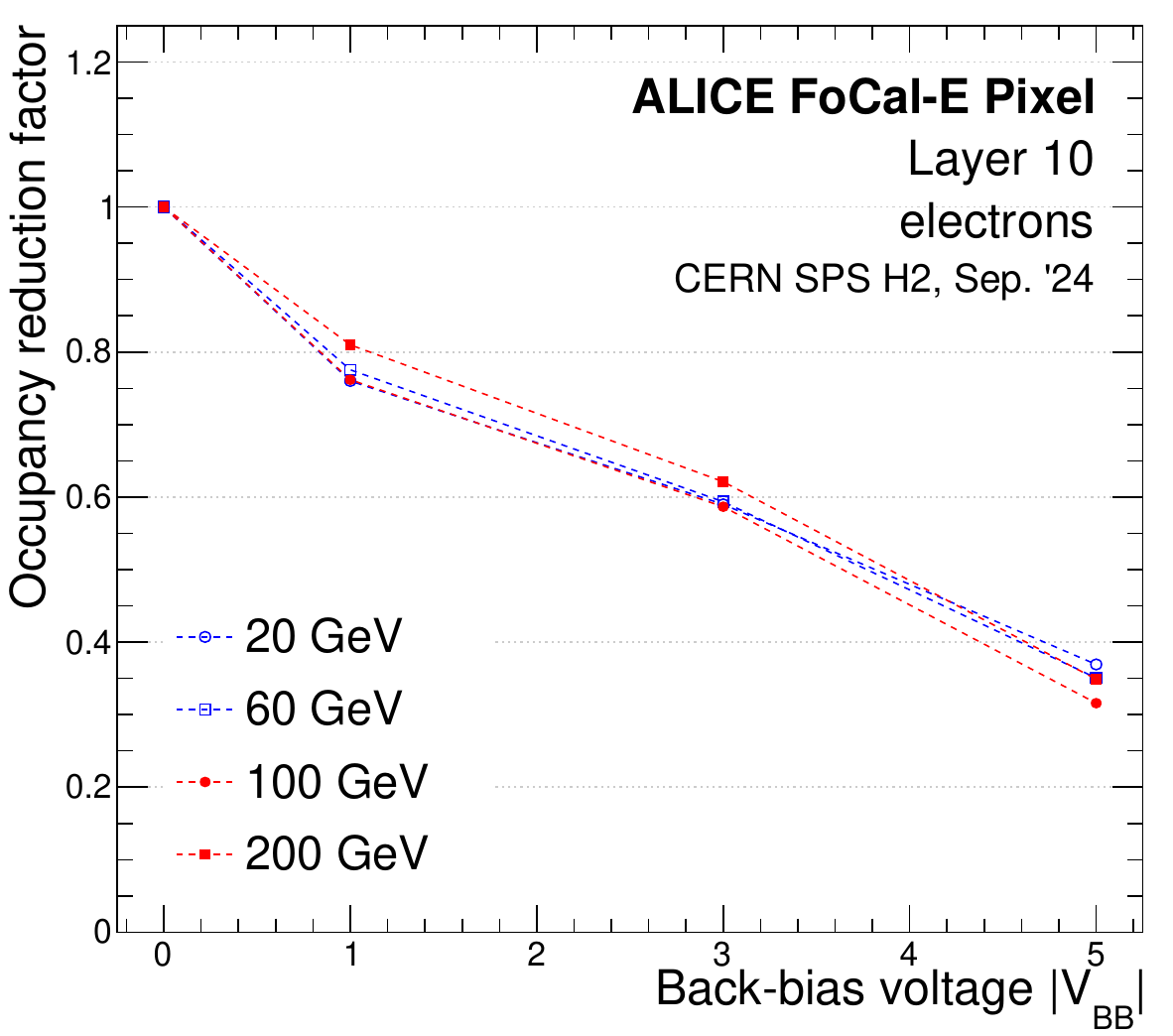}
    \end{minipage}
    \caption{(Top) Normalized hits in Layers 5 and 10 vs. \(|V_{BB}|\), highlighting a 
    reduction in recorded particle counts at higher back-bias voltages. (Bottom) Occupancy 
    reduction factor vs. \(|V_{BB}|\), showing further decreases at beam energies of 
    20--200\,GeV. Results from FoCal-E beam tests at CERN SPS H2.}
    \label{fig:FoCal-BB}
\end{figure}

As shown in \autoref{fig:FoCal-BB}, the normalized hit count in Layers~5 and 10 decreases 
notably with increasing \(|V_{BB}|\), indicating an effective reduction in occupancy. 
Similarly, the occupancy reduction factor improves across beam energies of 20--200\,GeV, 
confirming that back-biasing can significantly lower the overall hit density in high-energy 
electromagnetic showers.

\section{Exploring Optimization Options}
In the innermost region ($r<10$\,cm), high occupancy can cause ALPIDE BUSY Violations. 
Two complementary strategies—grid masking and a regional trigger—are under study to reduce occupancy and improve efficiency.

\subsection{Grid Masking}

\begin{figure}[h]
    \centering
    \includegraphics[width=0.85\linewidth]{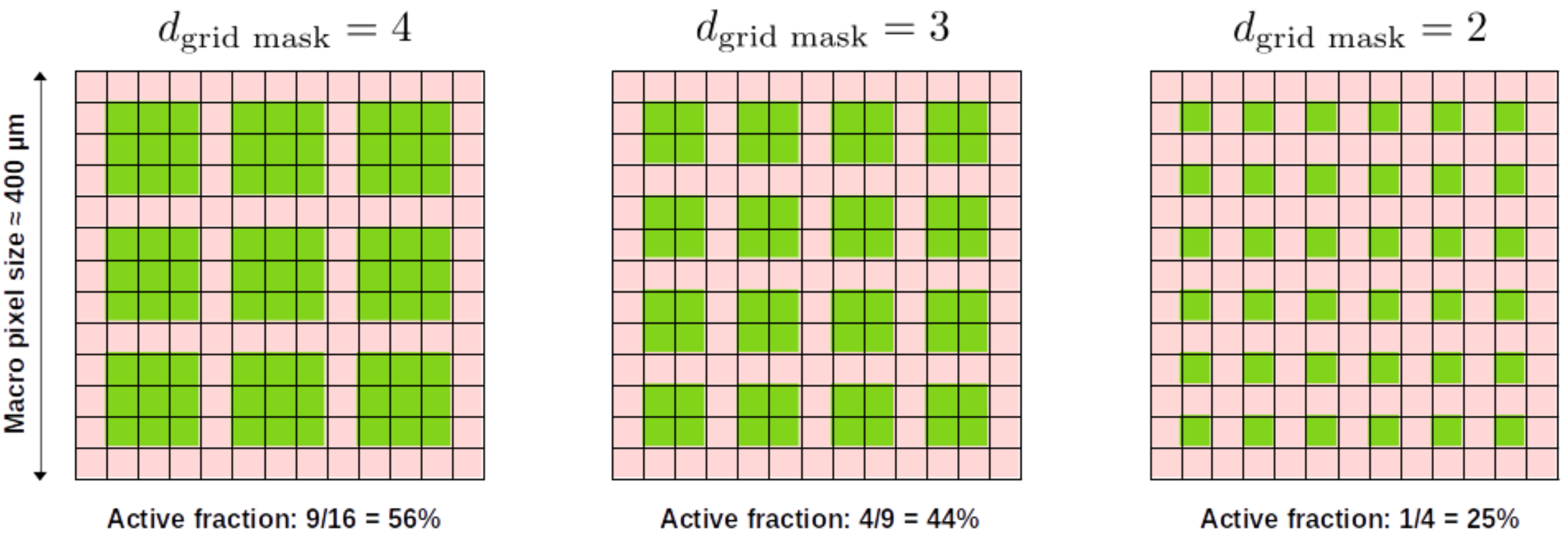}
    \caption{Simulated grid masks for the innermost chips at $r<10$\,cm, with a periodic pattern cell size $d_{\mathrm{grid\,mask}}$.}
    \label{fig:grid_masking}
\end{figure}

A straightforward approach is to mask a periodic fraction of pixels (see \autoref{fig:grid_masking}), thereby lowering occupancy in the innermost chips. 
For $d_{\mathrm{grid\,mask}}=4,3,2$, the active fraction is 56\%, 44\%, and 25\%, respectively. Although some fine-grained shower data is lost, $\pi^0$ separation mostly relies on larger macro-pixel cells, so the impact is tolerable. 
Ongoing studies focus on applying this mask only near the beam pipe, balancing occupancy reduction and physics performance.

\begin{figure}[h]
    \centering
    \includegraphics[width=0.9\linewidth]{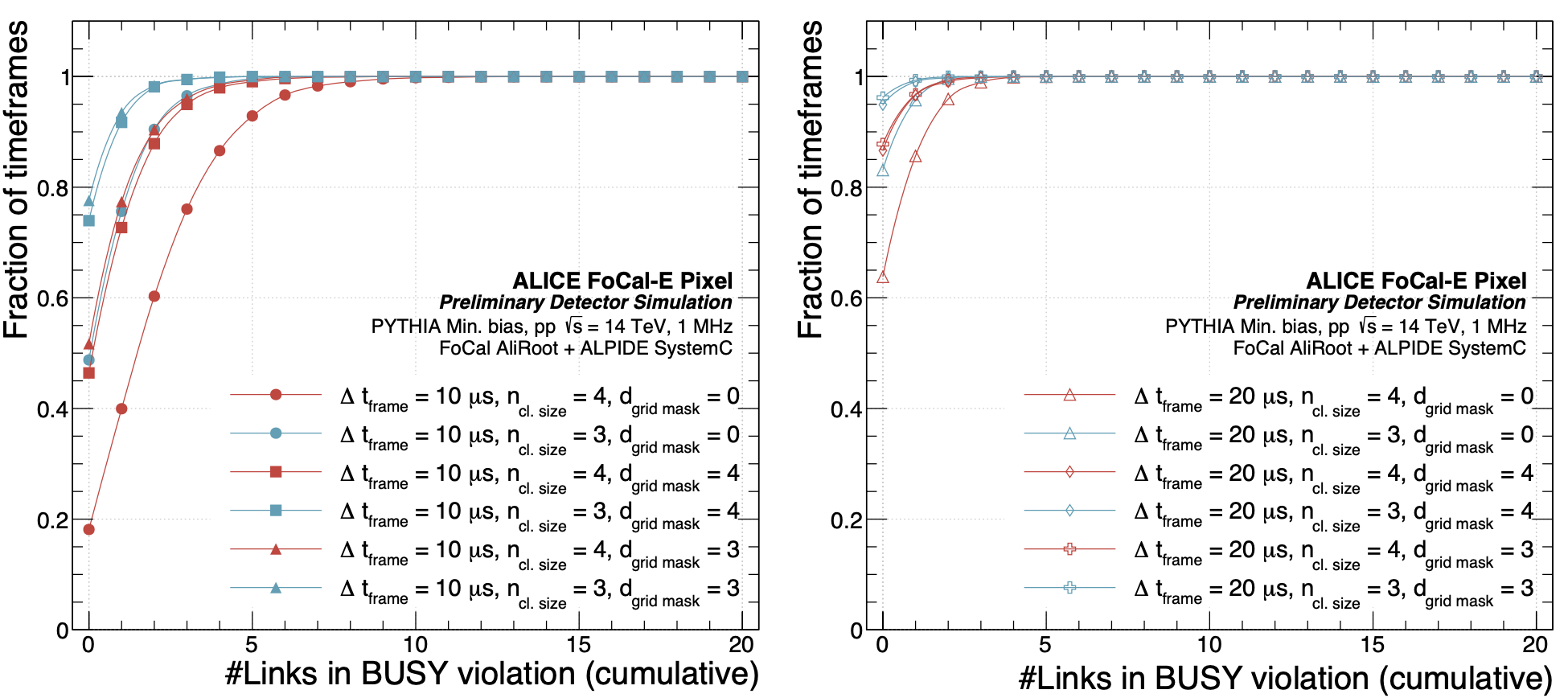}
    \caption{Fraction of frames with up to $n$ links in BUSY violation for pp collisions under different $d_{\mathrm{grid\,mask}}$ and $\Delta t_{\mathrm{frame}}$ \cite{CERN-LHCC-2024-004}.}
    \label{fig:BUSYV_link_vs_frac.tframe_grid}
\end{figure}

Under pp conditions (\autoref{fig:BUSYV_link_vs_frac.tframe_grid}) at $\Delta t_{\mathrm{frame}}=20\,\mu\mathrm{s}$ and $n_{\mathrm{cl.size}}=3$, the frame efficiency is about 85\% without grid masking, rising to nearly 95\% when $d_{\mathrm{grid\,mask}}=3$.

\subsection{Regional Trigger}

The Regional Trigger is implemented in the innermost FoCal-E area near the beam pipe to selectively activate ALPIDE sensors, thereby reducing occupancy and mitigating BUSY Violations. The pad wafer in the 4th pad layer, read out by HGCROC chips, monitors local energy deposition. Once the deposited energy exceeds a predefined threshold (e.g., 10--30\,MeV), only the corresponding ALPIDE sensors are triggered, significantly lowering the hit rate in the central zone.

\begin{figure}[h]
    \centering
    \begin{minipage}{0.43\linewidth}
        \centering
        \includegraphics[width=\linewidth]{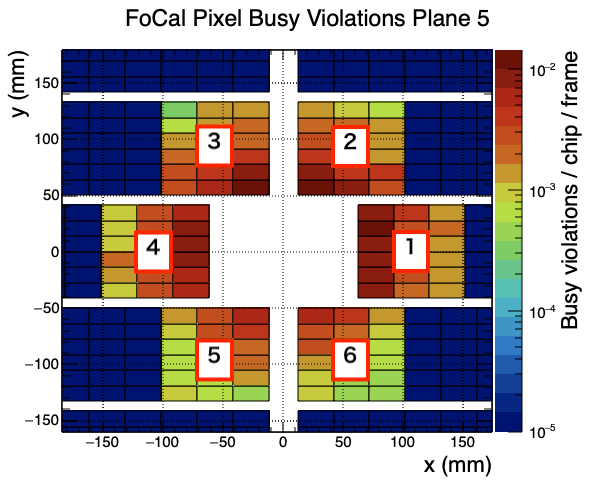}
    \end{minipage}
    \begin{minipage}{0.46\linewidth}
        \centering
        \includegraphics[width=\linewidth]{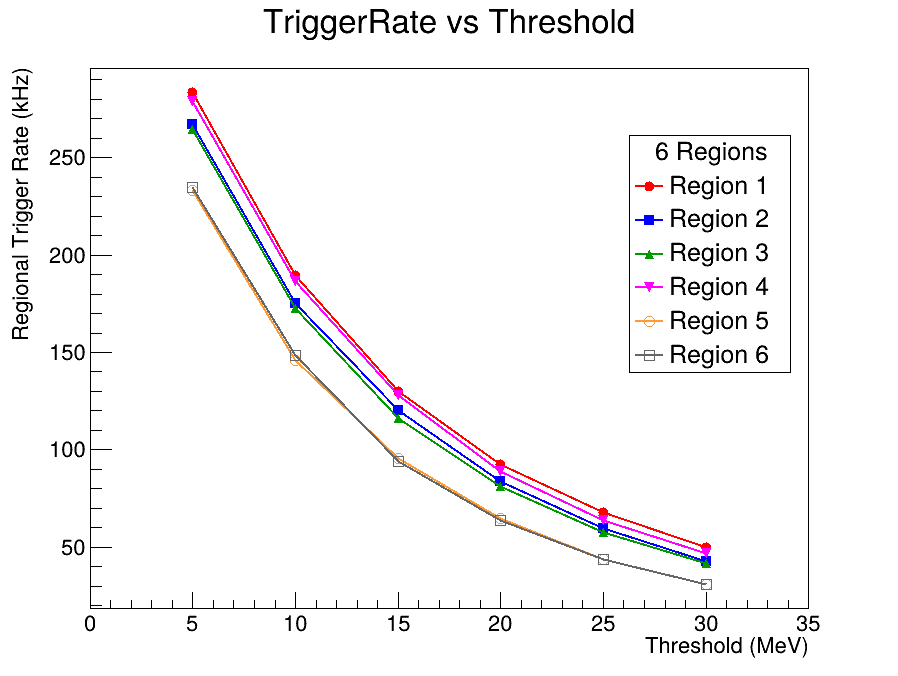}
    \end{minipage}
    \caption{(Left) Regional-trigger zones near the beam-pipe gap (color scale shows BUSY Violation rate at a 30\,MeV threshold). (Right) Trigger rate for different thresholds, based on SystemC simulations.}
    \label{fig:RegTrigger}
\end{figure}

\begin{figure}[h]
    \centering
    \includegraphics[width=0.9\linewidth]{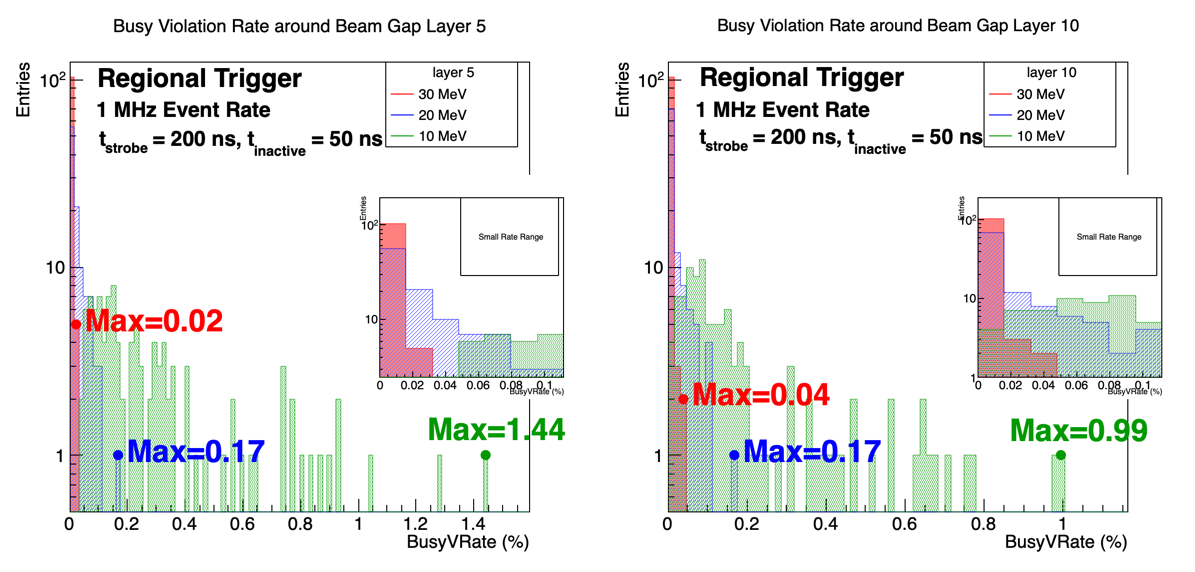}
    \caption{BUSY Violation rate at 10 MeV, 20 MeV, and 30 MeV thresholds, showing reduced occupancy in the central ALPIDE region.}
    \label{fig:BusyVRateComparison}
\end{figure}

As demonstrated in \autoref{fig:RegTrigger}, selective triggering effectively reduces the ALPIDE occupancy. Raising the threshold from 10\,MeV to 30\,MeV further decreases the trigger rate and BUSY violations down to about 0.04\%, greatly minimizing dead frames (\autoref{fig:BusyVRateComparison}). These results suggest that a regional trigger, possibly combined with grid masking, can be tuned to achieve optimal occupancy performance while preserving physics signals.

\section{Conclusion}
FoCal-E adopts ALPIDE-based pixel layers, demonstrating robust performance under high-multiplicity conditions. 
Data-rate analyses indicate the readout remains within acceptable throughput limits.
Beam test and SystemC simulation results confirm that back bias effectively suppresses pixel occupancy, thereby reducing dead-time frames. 
To further alleviate bottlenecks, grid masking and a regional trigger can bring maximum BUSY violations from a few percent down to 0.04\%, raising the fraction of valid frames above 95\%. 
These findings validate the feasibility of precise measurements of direct photons and neutral mesons in the forward region.

\section*{Acknowledges}
 This work was supported by the National Key R\&D Program of China (2022YFA1602103).

\bibliography{mybibfile}

\end{document}